\documentclass{article}
\usepackage{amssymb,latexsym}
\usepackage{amsmath}

\begin{document}
\title{Feynman's path integral to Ostrogradsky's Hamiltonian for Lagrangians with second derivatives}
\author{G.\ E.\ Hahne \thanks{Electronic address:  Gerhard.E.Hahne@nasa.gov}\\
M.\ S.\ 258-6, NASA/Ames Research Center\\
Moffett Field, California 94035 USA}
\maketitle

\centerline{PACS Numbers:  03.65.Ca, 03.65.Db}

\begin{abstract}
    A calculation is presented that shows that Feynman's path integral method
implies Ostrogradsky's Hamiltonian for nonsingular Lagrangians with second derivatives.  The procedure
employs the stationary phase approximation to obtain the limiting change of the wave function
per unit time.  By way of introduction, the method is applied anew to the case of nonsingular
Lagrangians with only first derivatives, but not necessarily quadratic in the velocities.
A byproduct of the calculation is an alternate derivation of the Legendre transformation for taking general classical
Lagrangians into Hamiltonians.
In both first and second derivative cases, the outcome contains precisely the classical Hamiltonian, which represents 
the so-called ``symbol'' of a (not necessarily Hermitean)
pseudodifferential operator acting on the wave function at an instant of time.
The derivation herein argues for a claim that Feynman's method starts with a classical Lagrangian
and ends with a classical Hamiltonian---nonclassical operator-ordering prescriptions in the passage
from classical to quantum Hamiltonians require external input and are generally
not inherent in Feynman's path integral formalism.

\end{abstract}

\section{Introduction} \label{S:sec1}

         Ostrogradsky (\cite{R:Ostrogradsky1}, \cite{R:Whittaker1}, \cite{R:Simon1}) gave a general formula for
transforming a Lagrangian formalism to a canonical, Hamiltonian formalism for general Lagrangians containing derivatives
up to $n^{\rm th}$ order for any chosen $n$.  We shall herein specialize to otherwise general Lagrangians with $n=1$ or $n=2$. 
Feynman \cite{R:Feynman1}, \cite{R:Brown1} proposed
a ``sum over classical paths'' as a means of obtaining the propagation kernel for Schr\"odinger's wave function from one time
to another later time.  The number of applications of Feynman's formalism, and of papers and books on the subject, is now huge---see
\cite{R:arXiv0}, \cite{R:Gutzwiller1}, \cite{R:arXiv1}, or any of the editions of \cite{R:Kleinert1}, for examples and references.  
A quick rundown of the basic quantum-mechanical method is given in {\em Wikipedia} \cite{R:Wiki1}.
As a warmup for the main subject, we now derive the Hamiltonian for a conventional Lagrangian 
(up to first derivatives only) by the method to be applied in Sec.~2
for Lagrangians with up to second derivatives.  

     The conventional nonrelativistic Schr\"odinger equation for
a Lagrangian $L_1$ (the subscript $1$ means that at most first derivatives appear)
comprising kinetic energy minus potential energy is derived \`a la Feynman in \cite{R:Feynman3}, p. 78, and in \cite{R:Derbes1}.
We shall follow a different procedure to this end, wherein the Legendre transformation appears as a consequence of the stationary phase approximation.
This approximation was used in a mathematical treatment \cite{R:Albeverio1} of oscillatory integrals in infinitely many dimensions,  this paper also
discussed the classical limit of 
$L=T-V$ quantum mechanics as an asymptotic expansion in powers of $\hbar^{-1}$---see \cite{R:Albeverio1}, Ch.~5. Schulman \cite{R:Schulman1}
made several uses of the stationary phase approximation, in particular, in obtaining a semiclassical approximation to
the propagator for nonzero time intervals (\cite{R:Schulman1}, Ch.~14) for conventional $L=T-V$ Lagrangians.

     Let time=$t$ be the independent variable, and one-dimensional position $x(t)$ be the dependent variable in the dynamics.
Dots, as in $\dot{x}(t)$, $\ddot{x}(t)$, and $\dddot{x}(t)$ stand for the first, second, and third derivatives,
respectively, of $x(t)$ with respect to time. We are given a Lagrangian $L_1(t,x(t),\dot{x}(t))$, a time interval $t_2-\Delta\leq t\leq t_2$,
and initial and final positions $x(t_2-\Delta)=x_1$ and $x(t_2)=x_2$.  For small $\Delta>0$, $x(t)$ is approximately linear and $\dot{x}(t)$ constant:
\begin{subequations}
\begin{gather}
x(t)\,\approx\,x_2-\frac{(t_2-t)}{\Delta}(x_2-x_1),\label{E1:gp1}\\
\dot{x}(t)\,\approx\, (x_2-x_1)/\Delta.\label{E1:gp2}
\end{gather}
\end{subequations}  
The action functional along this path is
\begin{equation}
S_1(t_2,x_2;t_2-\Delta, x_1)\,=\,\int_{t_2-\Delta}^{t_2}dt\,L_1(t,x(t),\dot{x}(t)).\label{E1:eq2}
\end{equation}
According to Feynman, we have for the Schr\"odinger wave function $\psi(t_2,x_2)$ in terms of the 
wave function $\psi(t_2-\Delta,x_1)$
\begin{equation}
\psi(t_2,x_2)\,\approx\,\int_{-\infty}^{+\infty}dx_1\sum_{\rm(paths)}\exp[(i/\hbar)S_1(t_2,x_2;t_2-\Delta,x_1)]N_1\psi(t_2-\Delta,x_1),\label{E1:eq3}
\end{equation}
where $N_1$ is a normalizing entity to be determined.  Following \cite{R:Feynman3}, p.~77, we presume that to calculate
the sum over kinematical paths, it is sufficient to take the single summand arising from the path \eqref{E1:gp1},
provided that $\Delta$ is sufficiently small.  (In \cite{R:Morette1}, the path integral is formulated in a manner that does not
involve Feynman's ``lattice approximation'';  the procedure for both $L_1$ and $L_2$ used herein is a lattice approximation with a one-cell lattice.) 

    We now suppose that a wave function at any time $t$ is given in terms of its wavenumber (i.e., momentum) $k$-space
representation $\phi(t,k)$ as follows:
\begin{equation}
\psi(t,x)\,=\,(2\pi)^{-1/2}\int_{-\infty}^{+\infty}dk \exp(ikx)\phi(t,k).\label{E1:eq4}
\end{equation}
Substituting \eqref{E1:eq4} into the rhs of \eqref{E1:eq3}, and interchanging the order of integration,  we find that
\begin{align}
\psi(t_2,x_2)\,&=\,(2\pi)^{-1/2}\int dk \phi(t_2-\Delta,k)\int dx_1 N_1\notag\\
&\ \  \ \times\exp\{(i/\hbar)[S_1(t_2,x_2;t_2-\Delta,x_1)+\hbar kx_1]\}.\label{E1:eq5}
\end{align}
We undertake to estimate the integral over $x_1$ on the rhs of \eqref{E1:eq5} by the stationary phase approximation \cite{R:Wiki2}:
we choose $(x_1)_{\rm s.p.}$ so that
\begin{equation}
0\,=\,\frac{\partial S_1}{\partial x_1}(t_2,x_2;t_2-\Delta,(x_1)_{\rm s.p.})+\hbar k.\label{E1:eq6}
\end{equation}
Given the $(x_1)_{\rm s.p.}$ such that \eqref{E1:eq6} is satisfied, a quadratic expansion of $S_1$ in $x_1$ at this point will enable a small $\Delta$
estimate of the normalization factor $N_1$ by an integral of Gaussian type---see Appendix A.

     We have
\begin{align}
S_1(t_2,x_2;t_2-\Delta,x_1)&\,=\,\int_{t_2-\Delta}^{t_2} dt L_1(t_2-(t_2-t), x_2-\dot{x}(t_2)(t_2-t),\dot{x}(t_2))\notag\\
&\approx\,\int _{t_2-\Delta}^{t_2} dt [L_1(t_2, x_2,\dot{x}(t_2))-\frac{\partial L_1}{\partial t}(t_2,x_2,\dot{x}(t_2))(t_2-t)\notag\\
&\ \ \ -\frac{\partial L_1}{\partial x}(t_2,x_2,\dot x(t_2))\dot{x}(t_2)(t_2-t)]\notag\\
&=\,L_1(t_2,x_2,\dot{x}(t_2))\Delta-\frac{\partial L_1}{\partial t}(t_2,x_2,\dot{x}(t_2))(\Delta^2/2)\notag\\
&\ \ \ -\frac{\partial L_1}{\partial x}(t_2,x_2,\dot{x}(t_2))\dot{x}(t_2)(\Delta^2/2).\label{E1:eq7}
\end{align}
The variables $t_2,\Delta,x_2,x_1$ are all independent, but from \eqref{E1:gp2} we have
\begin{equation}
\frac{\partial \dot{x}}{\partial x_1}(t_2)\,=\,-1/\Delta.\label{E1:eq8}
\end{equation}
Therefore, applying \eqref{E1:eq7} to \eqref{E1:eq6}, we have
\begin{equation}
0\,=\,-\frac{\partial L_1}{\partial \dot{x}}(t_2,x_2,\dot{x}(t_2))\,+\,\hbar k\,+\,O(\Delta)\label{E1:eq9}
\end{equation}
as $\Delta\to 0^+$, where the solution, if it exists and is
unique, to  
\eqref{E1:eq9} with the correction term omitted, determines the the stationary phase point $(x_1)_{\rm s.p.}$.
Note that we are assuming in obtaining \eqref{E1:eq9} 
that for any fixed $x_2$, $(x_2-(x_1)_{\rm s.p.})/\Delta=\dot{x}(t_2)$ tends to a finite value as $\Delta\to 0+$ (see \eqref{E1:eq11}, below).
This is plausible on physical grounds:  as $\Delta\to 0$, the stationary phase point should converge linearly
to $x_2$ in the limit.
Given that \eqref{E1:eq9} (to order $\Delta^0$) can be solved for
$\dot{x}(t_2)$ in terms of $t_2,x_2,\hbar k$: we infer that there is a unique function $F$ such that 
\begin{equation}
\dot{x}(t_2)\,=\, F(t_2,x_2,\hbar k),\label{E1:eq10}
\end{equation}
and 
\begin{equation}
(x_1)_{\rm s.p.}\,=\,x_2-\dot{x}(t_2)\Delta,=\,x_2-F(t_2,x_2,\hbar k)\Delta.\label{E1:eq11}
\end{equation}
We effect the Legendre transformation from $x,\dot{x}$ to canonical variables $x,\hbar k$, and from
Lagrangian to Hamiltonian, by defining the Hamiltonian $H_1(t_2,x_2,\hbar k)$ as
\begin{equation}
H_1(t_2,x_2,\hbar k)\,=\,\hbar k F(t_2,x_2,\hbar k)-L_1(t_2,x_2,F(t_2,x_2,\hbar k));\label{E1:eq12}
\end{equation}
we find that the exponential factor on the rhs of \eqref{E1:eq5} has the approximate form
\begin{equation}
\exp[i k x_2-(i\Delta/\hbar)H_1(t_2,x_2,\hbar k)+iO(x_1-(x_1)_{\rm s.p.})^2].\label{E1:eq13}
\end{equation}
We shall discuss the remaining integral over the exponential of the quadratic in $x_1$ in Appendix A;  we presume that the normalization entity $N_1$
cancels this integral, whereupon, to first order in $\Delta$,
\begin{equation}
\psi(t_2,x_2)\,=\,(2\pi)^{-1/2}\int dk [1-(i\Delta/\hbar)H_1(t_2,x_2,\hbar k)]\exp(ikx_2)\phi(t_2-\Delta,k).\label{E1:eq14}
\end{equation}
The limit $\Delta\to 0$ now entails
\begin{equation}
i\hbar\frac{\partial \psi}{\partial t}(t_2,x_2)\,=\,(2\pi)^{-1/2}\int dk H_1(t_2,x_2,\hbar k)\exp(ikx_2)\phi(t_2,k);\label{E1:eq15}
\end{equation}
in the latter form, $H_1$ is the classical Hamiltonian and acts as the so-called symbol of 
a pseudodifferential operator ($\Psi$DO) on (implicitly) $\psi(t_2,x_2)$---$\Psi$DO's
are defined in many places, as {\em Wikipedia} \cite{R:Wiki3}.  Note, however, that the $\Psi$DO engendered by $H_1(t_2,x_2,\hbar k)$
is not necessarily Hermitean:  for example, if $H_1$ contains a summand proportional to $x_2\hbar k$ the resulting $\Psi$DO is not Hermitean,
but the  summand $(x_2\hbar k-i\hbar/2)$ does yield a Hermitean operator. 
It is not possible, however, to get an imaginary contribution to the action function
from a one-classical-path approximation suitable for infinitesimal times.  

     Feynman (\cite{R:Feynman1}, Ch.~6) used a different approach to obtaining the Schr\"odinger equation from the classical path method, wherein
it was possible, at least in simple cases, to address
the problem of operator ordering.   A particular example was that of a nonrelativistic particle moving in an
electromagnetic field represented by a vector potential.  
Suppose, however, that we start with the classical Lagrangian
\begin{equation}
L_1\,=\,(m\dot{x}^2/2)(1+\alpha^2 x^2)^{-1},\label{E1:eq16}
\end{equation}
which yields the classical Hamiltonian 
\begin{equation}
H_1\,=\,p^2(1+\alpha^2x^2)/(2m).\label{E1:eq17}
\end{equation}
Suppose, furthermore, that the classical entity $x^2p^2$ is known 
to be represented by a certain Hermitean operator
\begin{equation}
[x^2p^2]_{\rm cl}\to[(x^2p^2+p^2x^2)/2+\hbar\beta(xp+px)/2+\hbar^2\gamma]_{\rm qm},\label{E1:eq18}
\end{equation}
where $\beta$ and $\gamma$ are real and dimensionless, and determined by nonclassical physics.
The stationary phase approximation
advocated herein is not able to resolve operator-ordering problems;  it seems fair to infer that Feynman's method, 
which begins with a classical Lagrangian, generally ends with a classical Hamiltonian rather than a quantum Hamiltonian.

     To extract Schr\"odinger quantum mechanics from \eqref{E1:eq15},   
operator ordering must be guided by physical and mathematical considerations external to the path integral formalism;
then, the argument $\hbar k$ in $H_1$ can be replaced by $(\hbar/i)(\partial/\partial x_2)$,
and the resulting $H_1$ operator taken outside the $k$-integral in \eqref{E1:eq15}.  That is, we can
obtain the usual Schr\"odinger
equation (with subscripts dropped on $t,x$, and appropriate operator ordering specified)
\begin{equation}
i\hbar\frac{\partial \psi}{\partial t}(t,x)\,=\,H_1(t,x,\frac{\hbar}{i}\frac{\partial}{\partial x})\psi(t,x).\label{E1:eq19}
\end{equation}

\section{Lagrangians with second derivatives} \label{S:sec2}

     In the case of a general Lagrangian $L_2(t,x,\dot{x},\ddot{x})$, the resulting variational principle yields a differential equation of
fourth order, so that the four dependent variables $x(t),\dot{x}(t),\ddot{x}(t),\dddot{x}(t)$ specify the state of a particle at time $t$.
Ostrogradsky (\cite{R:Ostrogradsky1},\cite{R:Whittaker1}) worked out a formalism that transforms from these variables to a canonical
formalism with two pairs of canonically conjugate dependent variables $(q_1(t),p_1(t)),(q_2(t),p_2(t))$ and a Hamiltonian
$H_2(t,q_1,q_2,p_1,p_2)$ such that the usual canonical equations of motion are satisfied.  Ostrogradsky's
procedure is as follows:
\begin{align}
q_1\,&=\,x,\label{E2:eq1}\\
q_2\,&=\,\dot{x},\label{E2:eq2}\\
p_1\,&=\,\frac{\partial L_2}{\partial \dot{x}}\,-\,\frac{d}{dt}\frac{\partial L_2}{\partial \ddot{x}},\notag\\
&=\,\frac{\partial L_2}{\partial \dot{x}}\,-\,\frac{\partial^2 L_2}{\partial t \partial\ddot{x}}
-\dot{x}\frac{\partial^2 L_2}{\partial x\partial \ddot{x}}\notag\\
&\ \ \ \ -\ddot{x}\frac{\partial^2 L_2}{\partial \dot{x}\partial \ddot{x}}
-\dddot{x}\frac{\partial^2 L_2}{\partial \ddot{x}\partial \ddot{x}}\label{E2:eq3}\\
p_2\,&=\,\frac{\partial L_2}{\partial \ddot{x}}.\label{E2:eq4}
\end{align}
We presume that these equations can be inverted uniquely:
$x(t),\dot{x}(t),\ddot{x}(t),\dddot{x}(t)$ are given in terms of $q_1,q_2,p_1,p_2$, as follows
\begin{align}
x\,&=\,q_1,\label{E2:eq5}\\
\dot{x}\,&=\,q_2,\label{E2:eq6}\\
\ddot{x}\,&=\,F_2(t,q_1,q_2,p_2)\label{E2:eq7}\\
\dddot{x}\,&=\,F_1(t,q_1,q_2,p_1,p_2).\label{E2:eq8}
\end{align}
The Hamiltonian takes the form
\begin{equation}
H_2(t,q_1,q_2,p_1,p_2)\,=\,-L_2(t,q_1,q_2,F_2)+p_1q_2+p_2F_2.\label{E2:eq9}
\end{equation}
Note that on account of \eqref{E2:eq1}, \eqref{E2:eq2}, and \eqref{E2:eq4}, $F_2$ does not depend on $p_1$, so that $p_1$ appears
only linearly in $H_2$:  this outcome gives rise to the instability (see, e.g., \cite{R:arXiv2}, and references given therein)
of the resulting classical and quantum mechanics, as the energy (=$H_2$) is unbounded both above and below.  

     We now proceed to show that the Feynman path integral yields a canonical formalism and Hamiltonian.
We shall consider the time interval $t_2-\Delta\leq t\leq t_2$ with $\Delta>0$ but small.  We also presume that
Feynman's sum over paths can be approximated by a single summand, that associated with the cubic curve
\begin{equation}
x(t)\,=\,x(t_2)-\dot{x}(t_2)(t_2-t)+\ddot{x}(t_2)(t_2-t)^2/2-\dddot{x}(t_2)(t_2-t)^3/6,\label{E2:eq10}
\end{equation}
where the curve is chosen so as to satisfy the four end-conditions
\begin{align}
x(t_2-\Delta)\,&=\,x_1,\label{E2:eq11}\\
\dot{x}(t_2-\Delta)\,&=\,\dot{x}_1,\label{E2:eq12}\\
x(t_2)\,&=\,x_2,\label{E2:eq13}\\
\dot{x}(t_2)\,&=\,\dot{x}_2.\label{E2:eq14}
\end{align}
We infer that
\begin{align}
\ddot{x}(t_2)\,&=\,-(6/\Delta^2)[(x_2-x_1-\Delta\dot{x}_2)+(\Delta/3)(\dot{x}_2-\dot{x}_1)]\label{E2:eq15}\\
\dddot{x}(t_2)\,&=\,-(12/\Delta^3)[(x_2-x_1-\Delta\dot{x}_2)+(\Delta/2)(\dot{x}_2-\dot{x}_1)]\label{E2:eq16}
\end{align}
We shall need the following for later applications:
\begin{align}
\frac{\partial}{\partial x_1}\ddot{x}(t_2)\,&=\,6/\Delta^2,\label{E2:eq17}\\
\frac{\partial}{\partial \dot{x}_1}\ddot{x}(t_2)\,&=\,2/\Delta,\label{E2:eq18}\\
\frac{\partial}{\partial x_1}\dddot{x}(t_2)\,&=\,12/\Delta^3,\label{E2:eq19}\\
\frac{\partial}{\partial \dot{x}_1}\dddot{x}(t_2)\,&=\,6/\Delta^2,\label{E2:eq20}
\end{align}

     Feynman's formula for the propagator in this case is
\begin{align}
\psi(t,x_2,\dot{x}_2)\,&=\, \int_{-\infty}^{+\infty}dx_1\int_{-\infty}^{+\infty}d\dot{x}_1\sum_{\rm (paths)}\notag\\
&\times\exp[(i/\hbar)S_2(t_2,x_2,\dot{x}_2;t_2-\Delta,x_1,\dot{x}_1)]N_2\psi(t_2-\Delta,x_1,\dot{x}_1).\label{E2:eq21}
\end{align}
In \eqref{E2:eq21}, $N_2$ is a normalizing entity discussed further in Appendix B, and we take the sum over paths to consist of the
single summand derived from the path \eqref{E2:eq10}:
\begin{align}
S_2(t_2,&x_2,\dot{x}_2;t_2-\Delta,x_1,\dot{x}_1)\,=\,\int_{t_2-\Delta}^{t_2}L(t,x(t),\dot{x}(t),\ddot{x}(t))dt\notag\\
&=\,\int_{t_2-\Delta}^{t_2}dt\notag\\
&\,\times L_2(t_2-(t_2-t),x_2-\dot{x}_2(t_2-t)+\ddot{x}(t_2)(t_2-t)^2/2-\dddot{x}(t_2)(t_2-t)^3/6,\notag\\
&\ \ \ \dot{x}_2-\ddot{x}(t_2)(t_2-t)+\dddot{x}(t_2)(t_2-t)^2/2,\ddot{x}(t_2)-\dddot{x}(t_2)(t_2-t))\notag\\
&\approx\,\int_{t_2-\Delta}^{t_2}\{L_2(t_2,x_2,\dot{x}_2,\ddot{x}(t_2))+\frac{\partial L_2}{\partial t}[-(t_2-t)]\notag\\
&\,+\frac{\partial L_2}{\partial x}[-\dot{x}_2(t_2-t)+\ddot{x}(t_2)(t_2-t)^2/2-\dddot{x}(t_2)(t_2-t)^3/6]\notag\\
&\,+\frac{\partial L_2}{\partial \dot{x}}[-\ddot{x}(t_2)(t_2-t)+\dddot{x}(t_2)(t_2-t)^2/2]\notag\\
&\,+\frac{\partial L_2}{\partial \ddot{x}}[-\dddot{x}(t_2)(t_2-t)]+\frac{\partial^2 L_2}{\partial t\partial t}(t_2-t)^2/2\notag\\
&\,+\frac{\partial^2 L_2}{\partial x\partial x}[-\dot{x}(t_2-t)+\ddot{x}(t_2)(t_2-t)^2/2-\dddot{x}(t_2)(t_2-t)^3/6]^2/2\notag\\
&\,+\frac{\partial^2 L_2}{\partial \dot{x}\partial\dot{x}}[-\ddot{x}(t_2)(t_2-t)+\dddot{x}(t_2)(t_2-t)^2/2]^2/2\notag\\
&\,+\frac{\partial^2 L_2}{\partial \ddot{x}\partial\ddot{x}}[-\dddot{x}(t_2)(t_2-t)]^2/2\notag\\
&\,+\frac{\partial^2 L_2}{\partial t\partial x}[-(t_2-t)][-\dot{x}(t_2-t)+\ddot{x}(t_2)(t_2-t)^2/2-\dddot{x}(t_2)(t_2-t)^3/6]\notag\\
&\,+\frac{\partial^2 L_2}{\partial t\partial\dot{x}}[-(t_2-t)][-\ddot{x}(t_2)(t_2-t)+\dddot{x}(t_2)(t_2-t)^2/2]\notag\\
&\,+\frac{\partial^2 L_2}{\partial t\partial\ddot{x}}(t_2-t)\dddot{x}(t_2)(t_2-t)\notag\\
&\,+\frac{\partial^2 L_2}{\partial x\partial \dot{x}}[-\dot{x}_2(t_2-t)+\ddot{x}(t_2)(t_2-t)^2/2-\dddot{x}(t_2)(t_2-t)^3/6]\notag\\
&\ \ \ \ \ \times[-\ddot{x}(t_2)(t_2-t)+\dddot{x}(t_2)(t_2-t)^2/2]\notag\\
&\,+\frac{\partial^2 L_2}{\partial x\partial \ddot{x}}[-\dot{x}(t_2-t)+\ddot{x}(t_2)(t_2-t)^2/2-\dddot{x}(t_2)(t_2-t)^3/6]\notag\\
&\ \times[-\dddot{x}(t_2)(t_2-t)]\notag\\
&\,+\frac{\partial^2 L_2}{\partial \dot{x}\partial\ddot{x}}[-\ddot{x}(t_2)(t_2-t)+\dddot{x}(t_2)(t_2-t)^2/2][-\dddot{x}(t_2)(t_2-t)]\}.
\label{E2:eq22}
\end{align}
We now carry out the integral over time, with the result (in the following rhs, we have discarded terms involving powers
of $\Delta$ greater than, or equal to, four, as these play a negligible role in the derivation due to \eqref{E2:eq19})
\begin{align}
S_2(t_2,&x_2,\dot{x}_2;t_2-\Delta,x_1,\dot{x}_1)\,=\,L_2(t,x_2,\dot{x}_2,\ddot{x}(t_2))\Delta
-\frac{\partial L_2}{\partial t}\Delta^2/2\notag\\
&\,+\frac{\partial L_2}{\partial x}[-\dot{x}_2\Delta^2/2+\ddot{x}(t_2)\Delta^3/6]
+\frac{\partial L_2}{\partial \dot{x}}[-\ddot{x}(t_2)\Delta^2/2+\dddot{x}(t_2)\Delta^3/6]\notag\\
&\,-\frac{\partial L_2}{\partial \ddot{x}}\dddot{x}(t_2)\Delta^2/2+\frac{\partial^2 L_2}{\partial t\partial t}[\Delta^3/6]
+\frac{\partial^2 L_2}{\partial x\partial x}\dot{x}_2^2\Delta^3/6\notag\\
&\,+\frac{\partial^2 L_2}{\partial \dot{x}\partial\dot{x}}\ddot{x}(t_2)^2\Delta^3/6
+\frac{\partial^2 L_2}{\partial \ddot{x}\partial\ddot{x}}\dddot{x}(t_2)^2\Delta^3/6\notag\\
&\,+\frac{\partial^2 L_2}{\partial t\partial x}\dot{x}_2\Delta^3/3
   +\frac{\partial^2 L_2}{\partial t\partial\dot{x}}\ddot{x}(t_2)\Delta^3/3\notag\\
&\,+\frac{\partial^2 L_2}{\partial t\partial\ddot{x}}\dddot{x}(t_2)\Delta^3/3
+\frac{\partial^2 L_2}{\partial x\partial \dot{x}}\dot{x}_2\ddot{x}(t_2)\Delta^3/3\notag\\
&\,+\frac{\partial^2 L_2}{\partial x\partial \ddot{x}}\dot{x}_2\dddot{x}(t_2)\Delta^3/3
+\frac{\partial^2 L_2}{\partial \dot{x}\partial\ddot{x}}\ddot{x}(t_2)\dddot{x}(t_2)\Delta^3/3\}.
\label{E2:eq23}
\end{align}

    We now define
\begin{align}
\Xi(k,k';t_2,x_2,\dot{x}_2;t_2-\Delta,x_1,\dot{x}_1)\ &=\ 
\hbar kx_1+\hbar k'\dot{x}_1
\notag\\
&\,+S_2(t_2,x_2,\dot{x}_2;t_2-\Delta,x_1,\dot{x}_1).
\label{E2:eq24}
\end{align}
In terms of the momentum representation of the initial wave function, Feynman's formula becomes
\begin{align}
\psi(t_2,x_2,\dot{x}_2)\,&=\,(2\pi)^{-1}\int dk\int dk'\phi(t_2-\Delta,k,k')N_2\int dx_1\int d\dot{x}_1
\notag\\
&\ \times\exp[(i/\hbar)\Xi(k,k';t_2,x_2,\dot{x}_2;t_2-\Delta,x_1,\dot{x}_1)].
\label{E2:eq25}
\end{align}
We estimate the inner double integral by the stationary phase approximation.   The stationary phase point
$((x_1)_{\rm s.p.},(\dot{x}_1)_{\rm s.p.})$, which we assume exists and is unique,
is determined by the two equations
\begin{equation}
\frac{\partial \Xi}{\partial x_1}(k,k';t_2,x_2,\dot{x}_2;t_2-\Delta,(x_1)_{\rm s.p.},(\dot{x}_1)_{\rm s.p.})\,=\,0,
\label{E2:eq26}
\end{equation}
\begin{equation}
\frac{\partial \Xi}{\partial \dot{x}_1}(k,k';t_2,x_2,\dot{x}_2;t_2-\Delta,(x_1)_{\rm s.p.},(\dot{x}_1)_{\rm s.p.})\,=\,0.
\label{E2:eq27}
\end{equation}
We now calculate the partial derivatives in \eqref{E2:eq26} and \eqref{E2:eq27}, using \eqref{E2:eq17}--\eqref{E2:eq20}
and \eqref{E2:eq23}.  We assume that for $(x_1,\dot{x}_1)$ kept at the ($\Delta$-dependent) stationary phase point as $\Delta$ decreases to zero
with $k,k',t_2,x_2,\dot{x}_2$ fixed, then $\ddot{x}(t_2)$ and $\dddot{x}(t_2)$ of \eqref{E2:eq15}, \eqref{E2:eq16} both tend to a finite limit;  
correspondingly, 
we neglect terms of order $\Delta^1$, $\Delta^2$, and $\Delta^3$ in the equations resulting from \eqref{E2:eq26} and \eqref{E2:eq27}
(more terms will be dropped later as appropriate).
We then infer from \eqref{E2:eq26} that
\begin{align}
0\,&=\,\hbar k+\frac{\partial L_2}{\partial\ddot{x}}(t_2,x_2,\dot{x}_2,\ddot{x}(t_2))(6/\Delta)+\frac{\partial L_2}{\partial \ddot{x}\partial t}(-3)
\notag\\
&\,+\frac{\partial L_2}{\partial \ddot{x}\partial x}(-3\dot{x}_2)+\frac{\partial L_2}{\partial \ddot{x}\partial\dot{x}}(-3\ddot{x}(t_2))\notag\\
&\,+\frac{\partial L_2}{\partial \dot{x}}(-3+2)+\frac{\partial L_2}{\partial \ddot{x}\partial\ddot{x}}(-3\dddot{x}(t_2))\notag\\
&\,+\frac{\partial L_2}{\partial\ddot{x}}(t_2,x_2,\dot{x}_2,\ddot{x}(t_2))(-6/\Delta)+\frac{\partial L_2}{\partial \ddot{x}\partial\ddot{x}}(4\dddot{x}(t_2))\notag\\
&\,+\frac{\partial L_2}{\partial \ddot{x}\partial t}(4)+\frac{\partial L_2}{\partial \ddot{x}\partial x}(4\dot{x}_2)\notag\\
&\,+\frac{\partial L_2}{\partial \ddot{x}\partial\dot{x}}(4\ddot{x}(t_2))+O(\Delta);\label{E2:eq28}
\end{align}
the terms in $\Delta^{-1}$ cancel, so that, consolidating results and taking $\Delta=0$, we obtain \eqref{E2:eq3}: 
\begin{equation}
\hbar k\,=\,\frac{\partial L_2}{\partial \dot{x}}\,-\,\frac{d}{dt}\frac{\partial L_2}{\partial \ddot{x}}.\label{E2:eq29}
\end{equation}
We also have from \eqref{E2:eq27}
\begin{equation}
0\,=\,\hbar k'+\frac{\partial L_2}{\partial \ddot{x}_2}(t_2,x_2,\dot{x}_2,\ddot{x}(t_2))(2-3)+O(\Delta),
\label{E2:eq30}
\end{equation}
that is, for $\Delta=0$,
\begin{equation}
\hbar k'\,=\,\frac{\partial L_2}{\partial \ddot{x}}(t_2,x_2,\dot{x}_2,\ddot{x}_2),\label{E2:eq31}
\end{equation}
which is \eqref{E2:eq4}.

     We solve \eqref{E2:eq31} for $\ddot{x}(t_2)$ in terms of $k',t_2,x_2,\dot{x}_2$ and then \eqref{E2:eq29}
for $\dddot{x}(t_2)$ in terms of $k,k',t_2,x_2,\dot{x}_2$:
\begin{align}
\ddot{x}(t_2)\,&=\,F_2(t_2,x_2,\dot{x}_2,\hbar k'),\label{E2:eq32}\\
\dddot{x}(t_2)\,&=\,F_1(t_2,x_2,\dot{x}_2,\hbar k,\hbar k'),\label{E2:eq33}
\end{align}
as in \eqref{E2:eq7}, \eqref{E2:eq8}.
We can now render self-consistent the previously assumed finite limits of \eqref{E2:eq15} and \eqref{E2:eq16} as $\Delta\to 0+$:
Turning these equations around, we find that
\begin{align}
x_2-(x_1)_{\rm s.p.}-\Delta\dot{x}_2\,&=\,-(\Delta^2/2)\ddot{x}(t_2)+(\Delta^3/6)\dddot{x}(t_2)\notag\\
&\,\approx-(\Delta^2/2)F_2+(\Delta^3/6)F_1,\label{E2:eq31a}\\
\dot{x}_2-(\dot{x}_1)_{\rm s.p.}\,&=\,\Delta\ddot{x}(t_2)-(\Delta^2/2)\dddot{x}(t_2)\approx \Delta F_2-(\Delta^2/2)F_1;\label{E2:eq31b}
\end{align}
therefore, if we substitute the rhs's of \eqref{E2:eq31a} and \eqref{E2:eq31b}, which give the
presumed limiting values of $\ddot{x}(t_2)$ and $\dddot{x}(t_2)$ as $\Delta\to 0$, into \eqref{E2:eq15} and {\eqref{E2:eq16}, we infer that
the rhs's of \eqref{E2:eq15} and \eqref{E2:eq16} tend to finite limits as $\Delta\to 0+$.

     At the stationary phase point, the first derivatives of $\Xi$ with respect to $x_1$ and $\dot{x}_1$ vanish,
and we estimate the integral over $x_1$ and $\dot{x}_1$ in \eqref{E2:eq25} using a quadratic approximation to $\Xi$
in the variables $(x_1-(x_1)_{\rm s.p.})$ and $(\dot{x}_1-(\dot{x}_1)_{\rm s.p.})$.  We shall consider 
the double integral involving the quadratic term,
which cancels the normalization factor, in Appendix B;  the zeroth order term to order $\Delta^1$ leads to the classical Hamiltonian, as follows.  
\begin{align}
\Xi(k,k';t_2,x_2,\dot{x}_2;&t_2-\Delta,(x_1)_{\rm s.p.},(\dot{x}_1)_{\rm s.p.})\,=\,
\hbar k (x_2-\dot{x}_2\Delta)+\hbar k'(\dot{x}_2-\ddot{x}(t_2)\Delta)\notag\\
&\,+\,L_2(t_2,x_2,\dot{x}_2,\ddot{x}(t_2))\Delta\notag\\
&=\,\hbar kx_2+\hbar k'\dot{x}_2-H_2(t_2,x_2,\dot{x}_2,F_1,F_2)\Delta,\label{E2:eq34}
\end{align}
where we have defined
\begin{equation}
H_2(t_2,x_2,\dot{x}_2,\hbar k,\hbar k')\,=\,-L_2(t_2,x_2,\dot{x}_2,\ddot{x}_2)+\hbar k\dot{x}_2+\hbar k'F_2.\label{E2:eq35}
\end{equation}
To first order in $\Delta$, therefore, \eqref{E2:eq25} comes down to
\begin{align}
\psi(t_2,x_2,\dot{x}_2)\,&=\,(2\pi)^{-1}\int dk\int dk'\exp[ikx_2+ik'\dot{x}_2]\notag\\
&\times[1-(i\Delta/\hbar)H_2(t_2,x_2,\dot{x}_2,\hbar k,\hbar k')]\phi(t_2-\Delta,k,k'),
\label{E_2:eq36}
\end{align}
that is, 
\begin{align}
i\hbar\frac{\partial \psi}{\partial t}(t_2,x_2,\dot{x}_2)\,&=\,
(2\pi)^{-1}\int dk\int dk'\exp[ikx_2+ik'\dot{x}_2]\notag\\
&\times H_2(t_2,x_2,\dot{x}_2,\hbar k,\hbar k')]\phi(t_2,k,k'),
\label{E2:eq37}
\end{align}
The latter manifests Ostrogradsky's classical Hamiltonian
$H_2$ as the symbol of a $\Psi$DO.  As in Sec.~1, \eqref{E1:eq15}, {\em et seq}., the  operator defined by $H_2$ is not necessarily Hermitean.
We must be guided here by considerations external to (at least) the one-classical-path approximation: that is, 
with appropriate operator ordering, we can express $\hbar k,\hbar k'$ in $H_2$
as differential operators and factor the resulting $H_2$ operator out of the integral, whereupon, dropping subscripts on $t,x,\dot{x}$, we have
\begin{equation}
i\hbar\frac{\partial\psi}{\partial t}(t,x,\dot{x})\,=\,H_2(t,x,\dot{x},\frac{\hbar}{i}\frac{\partial}{\partial x},
\frac{\hbar}{i}\frac{\partial}{\partial\dot{x}})\psi(t,x,\dot{x}).\label{E2:eq38}
\end{equation}
This is the Schr\"odinger equation (not necessarily uniquely) associated with Ostrogradsky's  Hamiltonian.

\begin{flushleft}
{\bf Appendix A:  Normalization with Lagrangian $L_1$}
\end{flushleft}

  In order to complete the stationary phase approximation and estimate $N_1$ in \eqref{E1:eq5}, 
we need to compute the second-order term in the exponent of \eqref{E1:eq5}
at the stationary phase point.  We use \eqref{E1:eq10} to find
\begin{align}
S_1(t_2,x_2;t_2-\Delta,x_1)\,&\approx\,S_1(t_2,x_2;t_2-\Delta,(x_1)_{\rm s.p.})\notag\\
&\,+\frac{\partial^2 L_1}{\partial\dot{x}\partial\dot{x}}(t_2,x_2,F(t_2,x_2,\hbar k))\notag\\
&\,\times(x_1-({x}_1)_{\rm s.p.})^2(2\Delta)^{-1}.
\tag{A1}
\end{align}
The normalization factor $N_1$ is therefore
\begin{equation}
N_1\,=\,[\frac{\partial^2 L_1}{\partial\dot{x}\partial\dot{x}}(t_2,x_2,F(t_2,x_2,\hbar k))/(2\pi i\hbar\Delta)]^{1/2},
\tag{A2}
\end{equation}
where the algebraic sign of the second derivative of $L_1$ must be accounted for.
Unless $L_1$ is quadratic in the velocity,
the entity $N_1$ is itself a nontrivial $\Psi$DO; hence its function in \eqref{E1:eq3} needs elaboration, which we shall not
pursue here.   We presume that the second
derivative of $L_1$ is slowly varying in $x_1$ as $\Delta\to 0$  
so that consistent with the stationary phase approximation \cite{R:Wiki2},
we have evaluated this entity at the stationary phase point. 

\begin{flushleft}
{\bf Appendix B:  Normalization with Lagrangian $L_2$}
\end{flushleft}

  Analogous to Appendix A, the normalization factor $N_2$ in \eqref{E2:eq25}  will be chosen to cancel the integral over 
$x_1,\dot{x}_1$ arising from the second-order
contribution in $(x_1-(x_1)_{\rm s.p.}),(\dot{x}_1-(\dot{x}_1)_{\rm s.p.})$ to the function $\Xi$ of \eqref{E2:eq24}.
We shall keep only the dominant term in a negative power of $\Delta$ in each of the derivatives of $S_2$ of \eqref{E2:eq23}:
the three terms in \eqref{E2:eq23} that provide such contributions comprise
\begin{align}
S_2(t_2,&x_2,\dot{x}_2;t_2-\Delta,x_1,\dot{x}_1)\,\approx\,L_2(t,x_2,\dot{x}_2,\ddot{x}(t_2))\Delta
-\frac{\partial L_2}{\partial \ddot{x}}\dddot{x}(t_2)\Delta^2/2\notag\\
&\,+\frac{\partial^2 L_2}{\partial \ddot{x}\partial\ddot{x}}\dddot{x}(t_2)^2\Delta^3/6.
\tag{B1}
\end{align}
The computations yield
\begin{align}
\frac{\partial^2 S_2}{\partial x_1\partial x_1}(t_2,x_2,\dot{x}_2;t_2-\Delta,(x_1)_{\rm s.p.},(\dot{x}_1)_{\rm s.p.})\,&\approx\,
\frac{12}{\Delta^3}\Bigl[\frac{\partial^2 L_2}{\partial\ddot{x}\partial\ddot{x}}\Bigr],\tag{B2}\\ 
\frac{\partial^2 S_2}{\partial x_1\partial \dot{x}_1}(t_2,x_2,\dot{x}_2;t_2-\Delta,(x_1)_{\rm s.p.},(\dot{x}_1)_{\rm s.p.})\,&\approx\,
-\frac{6}{\Delta^2}\Bigl[\frac{\partial^2 L_2}{\partial\ddot{x}\partial\ddot{x}}\Bigr],\tag{B3}\\ 
\frac{\partial^2 S_2}{\partial \dot{x}_1\partial \dot{x}_1}(t_2,x_2,\dot{x}_2;t_2-\Delta,(x_1)_{\rm s.p.},(\dot{x}_1)_{\rm s.p.})\,&\approx\,
\frac{4}{\Delta  }\Bigl[\frac{\partial^2 L_2}{\partial\ddot{x}\partial\ddot{x}}\Bigr],\tag{B4}
\end{align}
We define ${\mathcal D}$ as the determinant of the $2\times 2$ matrix of second derivatives:
\begin{align}
{\mathcal D}\,&=\,{\rm det}\left[\begin{matrix}
\partial^2 S_2/\partial x_1\partial x_1 & \partial^2 S_2/\partial x_1\partial \dot{x}_1\\
\partial^2 S_2/\partial \dot{x}_1\partial x_1 & \partial^2 S_2/\partial \dot{x}_1\partial \dot{x}_1
\end{matrix}\right],\tag{B5}\\
&=\,\frac{12}{\Delta^4}\Bigl[\frac{\partial^2 L_2}{\partial\ddot{x}\partial\ddot{x}}(t,x_2,\dot{x}_2,F_2(t,x_2,\dot{x}_2,\hbar k')\Bigr]^2.\tag{B6}
\end{align}
The normalization $N_2$ is the reciprocal of the integral of the phase:
\begin{equation}
N_2\ =\ {\mathcal D}^{1/2}/(2\pi i\hbar)
\ =\ (12)^{1/2}\Bigl[\frac{\partial^2 L_2}{\partial\ddot{x}\partial\ddot{x}}\Bigr]/(2\pi i\hbar\Delta^2),\tag{B7}
\end{equation}
where, as in (A2), the algebraic sign of the second derivative of $L_2$ must be taken into account.

\end{document}